# Sign Errors in "The Four Laws of Black Hole Mechanics"

Richard P. Behiel

March 26th, 2026

**Abstract.** In 1973, Bardeen, Carter, and Hawking published "The Four Laws of Black Hole Mechanics" [1], establishing the mathematical framework that would later be understood as the thermodynamics of black holes. Central to the paper is equation (33), which writes the variation of the total energy-momentum integral in terms of physically meaningful quantities: angular momentum, particle number, and entropy. Equation (33) feeds into the differential mass formula, equation (34), which is the first law of black hole mechanics. This note identifies two compensating sign errors in the BCH paper. The first error, demonstrated by a derivation from equation (32), is that equations (33) and (34) should carry minus signs rather than plus signs on the last two integrals, those involving the redshifted chemical potential $\bar{\mu}$ and the redshifted temperature $\bar{\theta}$. The second error is that the definitions of total particle number $N$ and total entropy $S$ given after equation (20) are missing minus signs that are required for these quantities to be positive. These two errors cancel, in that reversing the signs in the definitions of $N$ and $S$ to ensure positive quantities makes equations (33) and (34) correct.

*All conclusions of the BCH paper remain valid.* This note is intended merely as a guide for readers who, in working through the derivation step by step, might otherwise be puzzled by the sign discrepancies. Numbered equations refer to the BCH paper; lettered equations are introduced in this note.

### The Sign Error in Equations (33) and (34)

Equations (33) and (34) of the BCH paper both contain the same sign error: the $\bar{\mu}$ and $\bar{\theta}$ integrals in both equations should be negative, *if* the definitions of $N$ and $S$ in the paper are taken as written. In the next section, it will be shown that the signs of the definitions of $N$ and $S$ should be reversed, to ensure positive quantities, and this reversal will also resolve the sign error in the derivation of equation (33), and by extension (34). The conclusions of the BCH paper are thus still correct, once $N$ and $S$ are correctly defined.

To see that the definitions of $N$ and $S$ as written in the paper lead to the wrong signs on the $\bar{\mu}$ and $\bar{\theta}$ terms in equations (33) and (34), one can derive equation (33) from (32), and in the process encounter two unwanted minus signs. BCH equation (32) gives the variation of the energy-momentum tensor term in equation (28):

$$2\delta \int T_a^b K^a d\Sigma_b = -2 \int \Omega \delta\{T_a^b \widetilde{K}^a d\Sigma_b\} + 2\delta \int p K^a d\Sigma_a + 2 \int u^a \delta\left\{(\varepsilon + p)(-u^c u^d g_{cd})^{-1} u_a K^b d\Sigma_b\right\} \quad (32)$$

The paper then states that equation (32) can be written in the form of equation (33):

$$2\delta \int T_a^b K^a d\Sigma_b = \int T^{cd} h_{cd} K^a d\Sigma_a + 2 \int \Omega \delta dJ + 2 \int \bar{\mu} \delta dN + 2 \int \bar{\theta} \delta dS \quad (33)$$

where, as stated in the paper, $\delta dJ$, $\delta dN$, and $\delta dS$ are defined as:

$$\delta dJ = -\delta\{T_a^b \widetilde{K}^a d\Sigma_b\}, \qquad \delta dN = \delta\{n(-u_a u^a)^{-1/2} K^b d\Sigma_b\}, \qquad \delta dS = \delta\{s(-u_a u^a)^{-1/2} K^b d\Sigma_b\}$$

It will be shown now, by deriving (33) from (32), that the last two integrals in equation (33), involving $\bar{\mu}$ and $\bar{\theta}$, should be negative, if all equations and definitions in the BCH paper are taken exactly as written.

While deriving equation (33) from (32), right away it can be seen that the first integral in (32) is equal to the second integral in (33), because $-2\delta\{T_a^b \tilde{K}^a d\Sigma_b\} = 2\delta dJ$ by the definition of $\delta dJ$. We can therefore ignore those two integrals, and focus instead on the remaining terms from the RHS of (32) and (33), which we can label $(\widetilde{32})$ and $(\widetilde{33})$ respectively:

$$2\delta \int pK^a d\Sigma_a + 2 \int u^a \delta\{(\varepsilon + p)(-u^c u^d g_{cd})^{-1} u_a K^b d\Sigma_b\} \qquad (\widetilde{32})$$

$$\int T^{cd} h_{cd} K^a d\Sigma_a + 2 \int \bar{\mu}\delta dN + 2 \int \bar{\theta}\delta dS \qquad (\widetilde{33})$$

The question of whether (32) and (33) are equivalent is then reduced to the question of whether $(\widetilde{32})$ equals $(\widetilde{33})$. It will be seen that $(\widetilde{32})$ is not equal to $(\widetilde{33})$, but rather to a modified version of it in which the integrals involving $\bar{\mu}$ and $\bar{\theta}$ are negative. To see that this is the case, start with $(\widetilde{32})$ and recognize $-u^c u^d g_{cd}$ as $-u_c u^c$, then replace $\varepsilon + p$ with $\mu n + \theta s$ by equation (19), and replace $u_a$ with $(-u_c u^c)^{1/2} v_a$ by the definition of $v_a$ which is shown in BCH after equation (20). These substitutions transform $(\widetilde{32})$ into $(a)$:

$$2\delta \int pK^a d\Sigma_a + 2 \int u^a \delta\{(\mu n + \theta s)(-u_c u^c)^{-1/2} v_a K^b d\Sigma_b\} \qquad (a)$$

Note that $u_a$ is a timelike vector tangent to the flow lines of the fluid around the black hole, so the term $-u_c u^c$ is positive in the mostly-plus convention used in the paper.

Next, to go from equation $(a)$ to $(b)$, break up the $\mu n + \theta s$ in the second integral into two pieces, $\mu n$ and $\theta s$, then apply the product rule to each of those integrals, to split out $\mu$ and $\theta$ respectively:

$$2\delta \int pK^a d\Sigma_a + 2 \int u^a \delta\mu n(-u_c u^c)^{-1/2} v_a K^b d\Sigma_b + 2 \int \mu u^a \delta\{n(-u_c u^c)^{-1/2} v_a K^b d\Sigma_b\}$$

$$+ 2 \int u^a \delta\theta s(-u_c u^c)^{-1/2} v_a K^b d\Sigma_b + 2 \int \theta u^a \delta\{s(-u_c u^c)^{-1/2} v_a K^b d\Sigma_b\} \qquad (b)$$

By definition in BCH, $v^a = (-u_b u^b)^{-1/2} u^a$, so $u^a v_a = (-u_b u^b)^{1/2} v^a v_a = -(-u_b u^b)^{1/2}$ because $v_a$ is a timelike unit vector defined by normalizing $u_a$. We can therefore replace the $u^a v_a$ terms in the second and fourth integrals of $(b)$ with $-(-u_b u^b)^{1/2}$, which cancels out those $(-u_c u^c)^{-1/2}$ terms while giving each of those integrals a minus sign:

$$2\delta \int pK^a d\Sigma_a - 2 \int n\delta\mu K^b d\Sigma_b + 2 \int \mu u^a \delta\{n(-u_c u^c)^{-1/2} v_a K^b d\Sigma_b\}$$

$$- 2 \int s\delta\theta K^b d\Sigma_b + 2 \int \theta u^a \delta\{s(-u_c u^c)^{-1/2} v_a K^b d\Sigma_b\} \qquad (c)$$

The second and fourth integrals of $(c)$ can be combined into the second integral of $(d)$, by combining $n\delta\mu$ and $s\delta\theta$ into $\delta p$ using $n\delta\mu + s\delta\theta = \delta p$ as mentioned in the BCH paper in the sentence before equation (33):

$$2\delta \int pK^a d\Sigma_a - 2 \int \delta p K^b d\Sigma_b + 2 \int \mu u^a \delta\left\{n(-u_c u^c)^{-\frac{1}{2}} v_a K^b d\Sigma_b\right\} + 2 \int \theta u^a \delta\left\{s(-u_c u^c)^{-\frac{1}{2}} v_a K^b d\Sigma_b\right\} \qquad (d)$$

Next, we can break up the third and fourth integrals of $(d)$ with the product rule, splitting out $v_a$:

$$2\delta \int pK^a d\Sigma_a - 2\int \delta p K^b d\Sigma_b + 2\int \mu n(-u_c u^c)^{-1/2}K^b d\Sigma_b u^a \delta v_a + 2\int \mu u^a v_a \delta\{n(-u_c u^c)^{-1/2}K^b d\Sigma_b\}$$

$$+2\int \theta s(-u_c u^c)^{-1/2}K^b d\Sigma_b u^a \delta v_a + 2\int \theta u^a v_a \delta\{s(-u_c u^c)^{-1/2}K^b d\Sigma_b\} \quad (e)$$

$v^a$ is a unit timelike vector, so $v^a v_a = -1$ and therefore $u^a \delta v_a = \frac{1}{2}(-u_e u^e)^{1/2} v^c v^d h_{cd}$, where $h_{cd}$ is defined in the BCH paper as the variation of the metric tensor, $h_{ab} = \delta g_{ab}$. It should be noted that BCH writes $u^a \delta\left\{(-u^c u^d g_{cd})^{-1/2} u_a\right\} = \frac{1}{2} v^c v^d h_{cd}$ before equation (33), which is missing a factor of $(-u_e u^e)^{1/2}$. That appears to be a harmless typographical error which does not propagate into the derivation, as using that form of $u^a \delta v_a$ to derive (33) from (32) would spoil the $T^{cd} h_{cd}$ term.

Substitution of $u^a \delta v_a$ with $\frac{1}{2}(-u_e u^e)^{1/2} v^c v^d h_{cd}$ simplifies the third and fifth integrals in $(e)$:

$$2\delta \int pK^a d\Sigma_a - 2\int \delta p K^b d\Sigma_b + \int \mu n v^c v^d h_{cd} K^b d\Sigma_b + 2\int \mu u^a v_a \delta\{n(-u_c u^c)^{-1/2}K^b d\Sigma_b\}$$

$$+\int \theta s v^c v^d h_{cd} K^b d\Sigma_b + 2\int \theta u^a v_a \delta\{s(-u_c u^c)^{-1/2}K^b d\Sigma_b\} \quad (f)$$

The third and fifth integrals in $(f)$ can be combined into the third integral in equation $(g)$, using equation (19), $\mu n + \theta s = \varepsilon + p$. We can also substitute $u^a v_a = -(-u_e u^e)^{1/2}$, which was demonstrated previously, into the fourth and sixth integrals in $(f)$, which become the last two integrals in $(g)$:

$$2\delta \int pK^a d\Sigma_a - 2\int \delta p K^b d\Sigma_b + \int (\varepsilon + p) v^c v^d h_{cd} K^b d\Sigma_b$$

$$-2\int \mu(-u_e u^e)^{1/2}\delta\{n(-u_c u^c)^{-1/2}K^b d\Sigma_b\} - 2\int \theta(-u_e u^e)^{1/2}\delta\{s(-u_c u^c)^{-1/2}K^b d\Sigma_b\} \quad (g)$$

The last two integrals of $(g)$ simplify greatly by the recognition of $\delta dN = \delta\{n(-u_c u^c)^{-1/2}K^b d\Sigma_b\}$ and $\delta dS = \delta\{s(-u_c u^c)^{-1/2}K^b d\Sigma_b\}$, as well as $\bar{\mu} = (-u_e u^e)^{1/2}\mu$ and $\bar{\theta} = (-u_e u^e)^{1/2}\theta$. Those four equations are shown in BCH after equation (33). Making those substitutions, we have:

$$2\delta \int pK^a d\Sigma_a - 2\int \delta p K^b d\Sigma_b + \int (\varepsilon + p) v^c v^d h_{cd} K^b d\Sigma_b - 2\int \bar{\mu}\delta dN - 2\int \bar{\theta}\delta dS \quad (h)$$

In $(h)$, we already see those telltale minus signs on the last two integrals. All that remains now is to clean up the first few integrals in $(h)$, to show that $(h)$ is indeed equal to $(\overline{33})$, up to a difference in minus signs on the last two integrals. To that end, expand the first integral in $(h)$ with the product rule to split out $p$, using $\delta\{K^a d\Sigma_a\} = g^{cd} h_{cd} K^a d\Sigma_a / 2$ because $S$ is held fixed by BCH and $\delta K^a = 0$ by equation (21):

$$2\int \delta p K^a d\Sigma_a + \int p g^{cd} h_{cd} K^a d\Sigma_a - 2\int \delta p K^b d\Sigma_b + \int (\varepsilon + p) v^c v^d h_{cd} K^b d\Sigma_b - 2\int \bar{\mu}\delta dN - 2\int \bar{\theta}\delta dS \quad (i)$$

The first and third integrals in $(i)$ cancel, and the second and fourth integrals can be combined:

$$\int [(\varepsilon + p)v^c v^d + pg^{cd}]h_{cd}K^b d\Sigma_b - 2\int \bar{\mu}\delta dN - 2\int \bar{\theta}\delta dS \qquad (j)$$

Recognizing $T^{cd} = (\varepsilon + p)v^c v^d + pg^{cd}$ from equation (20), we can write the first integral in $(j)$ as:

$$\int T^{cd}h_{cd}K^b d\Sigma_b - 2\int \bar{\mu}\delta dN - 2\int \bar{\theta}\delta dS \qquad (k)$$

Equation $(k)$ is nearly identical to equation $(\widetilde{33})$, except that the last two integrals are negative. This implies that, if the definitions of $N$ and $S$ are taken as written in BCH, equation (33) of the BCH paper would read:

$$2\delta \int T_a^b K^a d\Sigma_b = \int T^{cd}h_{cd}K^a d\Sigma_a + 2\int \Omega \delta dJ - 2\int \bar{\mu}\delta dN - 2\int \bar{\theta}\delta dS \qquad (l)$$

Moreover, those minus signs would propagate directly into the differential mass formula, equation (34), which inherits the same sign structure as equation (33). Thus, with the definitions of $N$ and $S$ as written in the BCH paper, both equations (33) and (34) contain sign errors on the $\bar{\mu}$ and $\bar{\theta}$ terms. However, those same definitions of $N$ and $S$ are themselves wrong by a sign, as will be shown in the next section. Equation $(l)$, which follows from (32), leads to a version of the first law of black hole mechanics in which negative $N$ and $S$ contribute negatively to the mass of the black hole. This double negative is of no physical significance. When the signs of $N$ and $S$ are corrected, $\delta dN$ and $\delta dS$ each acquire a minus sign, cancelling the two errant minus signs that would otherwise appear between equations $(g)$ and $(h)$, and equations (33) and (34) then both become correct as written in BCH.

### The Sign Error in the Definitions of $N$ and $S$

The sign error on the $\bar{\mu}$ and $\bar{\theta}$ terms in equations (33) and (34) is resolved by a complementary error in the definitions of $N$ and $S$. Because the sign error affects only the $\bar{\mu}$ and $\bar{\theta}$ integrals in equations (33) and (34), leaving the $\delta dJ$ term intact, it is unlikely to have originated from a sign error in equation (32). The complementary error therefore must pertain to either the signs of $\delta dN$ and $\delta dS$, or of $\bar{\mu}$ and $\bar{\theta}$, since reversing the sign of either pair would resolve the error. However, reversing the signs of $\bar{\mu}$ and $\bar{\theta}$ would introduce a negative redshift factor, which is unphysical. The only remaining option is that the definitions of $\delta dN$ and $\delta dS$ require sign reversal, via redefinition of the quantities $N$ and $S$. This sign reversal must, of course, be motivated by argumentation which stands independently of the desire to fix equations (33) and (34).

Consider the definitions of $J$, $S$, and $N$ given in the BCH paper after equation (20):

$$J = -\int T^a_b \widetilde{K}^b d\Sigma_a \qquad S = \int sv^a d\Sigma_a \qquad N = \int nv^a d\Sigma_a$$

The definition of $J$ carries an explicit minus sign, while the definitions of $S$ and $N$ do not. This asymmetry is inconsistent with the sign conventions required by the volume element. In BCH's mostly-plus signature, the future-directed unit normal to a spacelike hypersurface has a positive contravariant time component, so its covariant time component is negative, and therefore $d\Sigma_0 < 0$. This sign is fixed by the requirement that equation (6) yields $M > 0$ for positive energy density. The definition of $J$ accounts for $d\Sigma_0 < 0$, but the definitions of $S$ and $N$ do not.

To confirm that the definition of $J$ has the correct sign, consider slowly rotating dust with prograde angular momentum, $T_\phi^0 > 0$, so $T_\phi^0 d\Sigma_0 < 0$. The explicit minus sign in the definition of $J$ ensures that prograde rotation gives $J > 0$. On the other hand, the definitions of $N$ and $S$ do not include an explicit minus sign, and therefore take on unphysical negative values for positive $n$ and $s$, respectively. For concreteness, consider static dust with positive particle density $n > 0$ and four-velocity $v^a = [1,0,0,0]$ in Minkowski spacetime, the asymptotic limit of BCH:

$$N = \int nv^a d\Sigma_a = \int nv^0 d\Sigma_0 < 0$$

The BCH definition of $N$ yields negative particle number for positive particle density $n$, which is unphysical. The same argument applies to $S$ and $s$. The corrected definitions of $N$ and $S$ are therefore:

$$N = -\int nv^a d\Sigma_a \qquad S = -\int sv^a d\Sigma_a$$

The corresponding differentials then become $\delta dN = -\delta dN_{BCH}$ and $\delta dS = -\delta dS_{BCH}$, which resolves the sign error in equations (33) and (34).

## Conclusion

In summary, the BCH paper contains a sign error pertaining to the $\bar{\mu}$ and $\bar{\theta}$ integrals in equations (33) and (34), which is contingent upon another sign error in the definitions of $N$ and $S$. If the definitions of $N$ and $S$ are taken as written in the BCH paper, then equations (33) and (34) are incorrect, because the $\bar{\mu}$ and $\bar{\theta}$ terms should be negative. If, however, the signs are reversed in both the definitions of $N$ and $S$, then $N$ and $S$ become positive, and equations (33) and (34) become correct as written in BCH. Because these complementary sign errors cancel out, they have no physical consequences, and all conclusions of the BCH paper remain valid.

It should also be noted that the BCH paper contains two additional typographical errors in its intermediate expressions: the identity preceding equation (26), written as $(\delta l^a)_{;b} l^b + \delta l_a l^a_{;b}$ should read $(\delta l_a)_{;b} l^b + \delta l_b l^b_{;a}$ to satisfy the definition of the Lie derivative of a covector and to have consistent free indices on both terms; and, the equation stated before equation (33), $u^a \delta \left\{ \left( -u^c u^d g_{cd} \right)^{-1/2} u_a \right\} = \frac{1}{2} v^c v^d h_{cd}$, is missing a factor of $(-u_e u^e)^{1/2}$ on the right-hand side. Those two typographical errors have not affected the logical structure of the paper, as the correct versions of both are necessarily used when solving for $\delta \kappa$ and deriving equation (33), respectively.

## Acknowledgements

Thanks to Nirmalya Kajuri (IIT Mandi) for independently verifying the results of this note, and for helpful comments on the manuscript.